# EEG Machine Learning for Analysis of Mild Traumatic Brain Injury: A survey

Weiqing Gu, Ryan Chang, Bohan Yang

gu@dasion.ai

by93@cornell.edu

rchang05@usc.edu

## Abstract:

Mild Traumatic Brain Injury (mTBI) is a common brain injury and affects a diverse group of people: soldiers, constructors, athletes, drivers, children, elders, and nearly everyone. Thus, having a well-established, fast, cheap, and accurate classification method is crucial for the well-being of people around the globe. Luckily, using Machine Learning (ML) on electroencephalography (EEG) data shows promising results. This survey analyzed the most cutting-edge articles from 2017 to the present. The articles were searched from the Google Scholar database and went through an elimination process based on our criteria. We reviewed, summarized, and compared the fourteen most cutting-edge machine learning research papers for predicting and classifying mTBI in terms of 1) EEG data types, 2) data preprocessing methods, 3) machine learning feature representations, 4) feature extraction methods, and 5) machine learning classifiers and predictions. The most common EEG data type was human resting-state EEG, with most studies using filters to clean the data. The power spectral, especially alpha and theta power, was the most prevalent feature. The other non-power spectral features, such as entropy, also show their great potential. The Fourier transform is the most common feature extraction method while using neural networks as automatic feature extraction generally returns a high accuracy result. Lastly, Support Vector Machine (SVM) was our survey's most common ML classifier due to its lower computational complexity and solid mathematical theoretical basis. The purpose of this study was to collect and explore a sparsely populated sector of ML, and we hope that our survey has shined some light on the inherent trends, advantages, disadvantages, and preferences of the current state of machine learning-based EEG analysis for mTBI.

# I. Introduction

## 1.1 Introduction

Concussion or mild Traumatic Brain Injury (mTBI) is a prevalent symptom, with more than 3 million cases in the US per year [14]. It is a traumatically induced physiological disruption of brain function by an external force, such as a blow or violent shaking, that can disrupt memory, movement, sensation, and sleep-wake rhythm [11]. Military events, sports, and motor collisions are the most common causes of mTBI. mTBI mostly causes problems with headaches, fatigue, and memory loss. It can cause a person to develop chronic migraines, vomit, and cause irritability. A person who suffers from an mTBI is more susceptible to bouts of depression. Furthermore, over 70-90% of traumatic brain injury is mild. Because mTBI is so common and minor but greatly impacts a person's cognitive and physical ability, it is often called a "silent epidemic." In extreme cases, a dangerous blood clot that crowds the brain against the skull can develop [13].

Brain Injury can be classified using Glasgow Coma Scale (GCS) into three levels of severity: mild, moderate, or severe. GCS scores the severity based on patients' eye-opening, verbal, and motor responses [12]. If the score is between the interval of 3-8, it is a severe TBI. If the score is between 9-13, it is a moderate TBI. If the score is between 14-15, it is a mild TBI. However, GCS is not perfect: it is a qualitative test. Therefore, it is highly subjective, which can cause human bias. Furthermore, the eye-opening response might be unattainable during eye injury [59].

Magnetic Resonance Imaging (MRI) and Computed Tomography (CT) are standard clinical tools to detect brain dysfunctions in the medical field. However, they generally fail to reveal any persistent abnormalities in mTBI patients [56]. Furthermore, researchers found that scans can disturb sleep-wake rhythm and delay the recovery of patients [3]. There is also a concern that rapid radiated scans can harm patients' health [4]. Many electrophysiological methods, such as magnetoencephalography (MEG), show promising results. However, they are time-consuming, expensive, and require specialized operators. Due to limited resources, they can cause the delay of performing scans and expose patients to the potential risk of delayed treatment [2].

Among these potential concerns, an alternative solution is proposed: applying machine learning on Electroencephalography (EEG) data. Although conventional EEG analysis shows limited evidence, quantitative EEG (qEEG) with machine learning techniques seems promising [56]. In contrast to CT, MRI, and MEG, EEG is faster, cheaper, and portable. In addition, while EEG does have lower spatial resolutions (how accurately the measured activity is localized within the brain), it has a high temporal resolution (how closely the measured activity corresponds to the timing of the actual neuronal activity) [5,6].

While EEG has proven to be an effective tool for many fields, it continues to be held back by a few limitations that make it challenging to analyze and process it effectively. Due to the poor signal-to-noise ratio of EEG, a good signal-to-noise ratio is necessary to analyze large subject-specific, inter-, and intra-trial variability; hence, a detailed analysis is needed [19]. Moreover, EEG data possesses high inter-subject variability, which further limits its effectiveness of EEG data. This problem stems from the differences in physiology between individuals [18]. Moreover, it is difficult to differentiate between activities occurring at closely adjacent locations [20]. This occurs due to the possibility of particularly strong electrical activity being picked up by several neighboring electrodes [21]. However, today, there are more advanced methods, machine learning, for the estimates of EEG data from a single source.

Quantitative Electroencephalogram, or qEEG, is a diagnostic tool that measures electrical activity in the form of brain wave patterns. Sometimes it is referred to as "brain mapping." Brain waves are the rhythmic electrical impulses in the brain when the neurons communicate with one another. Neurons help transport an individual's behavior, emotions, and thoughts within the brain. Brain waves can reveal crucial information about one's general brain function including but not limited to stress levels, thought patterns, and emotions. qEEG can identify brain wave patterns that are associated with a wide range of symptoms such as impulsivity, anxiety, and more. Similar to EEG, qEEG is noninvasive, painless, and safe to use for patients of all ages.

However, while EEG records electrical activity that is representative of the underlying cortical brain activity, qEEG applies sophisticated mathematical and statistical analysis to the recorded electrical activity and compares them to control subjects. qEEG has had much success in a variety of diagnostic and informative settings. Similarly, qEEG is a "relatively easy to apply, cost-effective method among many electrophysiologic and functional brain imaging techniques used to assess individuals for prognosis and determination of the most suitable treatment." Its temporal resolution and "dimensional approach to the symptomatology of psychiatric disorders" are both invaluable advantages for qEEG.

## 1.2 Advantages and Disadvantages of EEG

There are several advantages of EEG over other brain scan technologies, such as Magnetic Reasoning Imaging (MRI), Magnetoencephalography (MEG), or Computed Tomography (CT). In general, EEG is cheaper, easier to use, faster, and portable. The average cost of an EEG scan without insurance ranges from $200 - $950. While the average cost without insurance for MRI is from $700 - $3500, and for CT, this number can be greater than $4000. [2] On average, the mass of an EEG device is lower than one kilogram. This makes EEG devices portable to use [see the device table]. Most importantly, it has a high temporal resolution, which can measure brain

activity on the order of milliseconds. This unique trait allows EEG to detect small disturbances in brain activity, which gives it the potential to diagnose diseases at an early stage or detect diseases that have weaker symptoms. Lastly, it is also non-invasive. This means that it has little or nearly no effect on patients; in contrast with other techniques: CT uses radiation, and MRI uses a strong magnet. These methods might disturb patients' sleep rhythm and delay their recovery.

After saying all these advantages, EEG also has some limitations, waiting for future researchers to conquer. EEG is easily contaminated by noises and other electrical devices during the recording. As up-to-date, there is no technique to decontaminate data. If a piece of EEG data is contaminated by noises, there is no way to eliminate it, just as people cannot turn a boiled egg back to raw. Second, EEG is limited to recording the structures deeper than the cortex. This makes EEG impossible to record and detect those diseases that occur deeper than the cortex. Lastly, one of the biggest limitations is that it has a poor spatial resolution. Because EEG monitors activity in large groups of neurons, it is difficult to pinpoint activity to a precise location in the brain. This leads to another critical drawback, that EEG classification is not specific enough. By performing EEG, researchers can easily figure out whether there are any abnormalities in the brain by comparing it to the normative database. However, there is not a clear cut between different brain diseases. The EEG of mTBI can look similar to the EEG of Alzheimer's disease, and the EEG of Alzheimer's disease can look similar to the EEG of Parkinson's disease. As researchers are trying to diagnose different diseases, such as mTBI or Alzheimer's disease, their results are often very high when they perform binary classification, healthy control vs. abnormal.

## 1.3 The Emerge of the qEEG

Quantitative Electroencephalogram, or qEEG, is a diagnostic tool that measures electrical activity in the form of brain wave patterns. Sometimes it is referred to as "brain mapping." Brain waves are the rhythmic electrical impulses in the brain when the neurons communicate with one another. Neurons help transport an individual's behavior, emotions, and thoughts within the brain. Brain waves can reveal crucial information about one's general brain function including but not limited to stress levels, thought patterns, and emotions. qEEG can identify brain wave patterns that are associated with a wide range of symptoms such as impulsivity, anxiety, and more. Similar to EEG, qEEG is noninvasive, painless, and safe to use for patients of all ages.
However, while EEG records electrical activity that is representative of the underlying cortical brain activity, qEEG applies sophisticated mathematical and statistical analysis to the recorded electrical activity and compares them to control subjects. qEEG has had much success in a variety of diagnostic and informative settings. Similarly, qEEG is a "relatively easy to apply, cost-effective method among many electrophysiologic and functional brain imaging techniques used to assess individuals for prognosis and determination of the most suitable treatment." Its temporal resolution and "dimensional approach to the symptomatology of psychiatric disorders" are both invaluable advantages for qEEG.

Following this introduction, we will talk about our literature review methodology in Section II. Then, in Section III, we describe each paper we found during the literature review in detail: from EEG data, data filtering methods, machine learning features, and feature extractions, to machine learning methods. Lastly, we will give a discussion and conclusion in Sections IV & V

# II. Methodology

## 2.1 Search Report

This survey uses Google Scholars databases. We first used a combination of keywords as input to the search engine: "electroencephalography," "mild Traumatic Brain Injury," "Machine learning," "Artificial Intelligence," "classification," and "accuracy." Next, to ensure the survey is accurate and up-to-date, we set the time range as "2017 or later." Next, the sorting is set as "Sort by relevance." As a result, 360 articles are generated and go through a selection process to eliminate any disqualifiers that do not align with the criteria.

In the first round of elimination, we glance through the title and the abstract to eliminate any article that

1. is not in English,
2. does not mention electroencephalography or EEG,
3. does not mention mTBI, mild Traumatic Brain Injury, or concussion,
4. does not mention any types of machine learning methods,
5. are not scholarly research papers,
6. does not have the full-text version publicly available,
7. are repeated titles or repeated contents with a slightly different title.

In the first elimination process, 318 articles were eliminated from our survey. In the second round of elimination, we carefully examine the content of each article and eliminate the article if

1. electroencephalography or EEG is not its main content,
2. mild Traumatic Brain Injury or mTBI is not its main content,
3. no classification accuracy is reported.

In the end, 15 articles remain after the elimination process and are included in our survey.

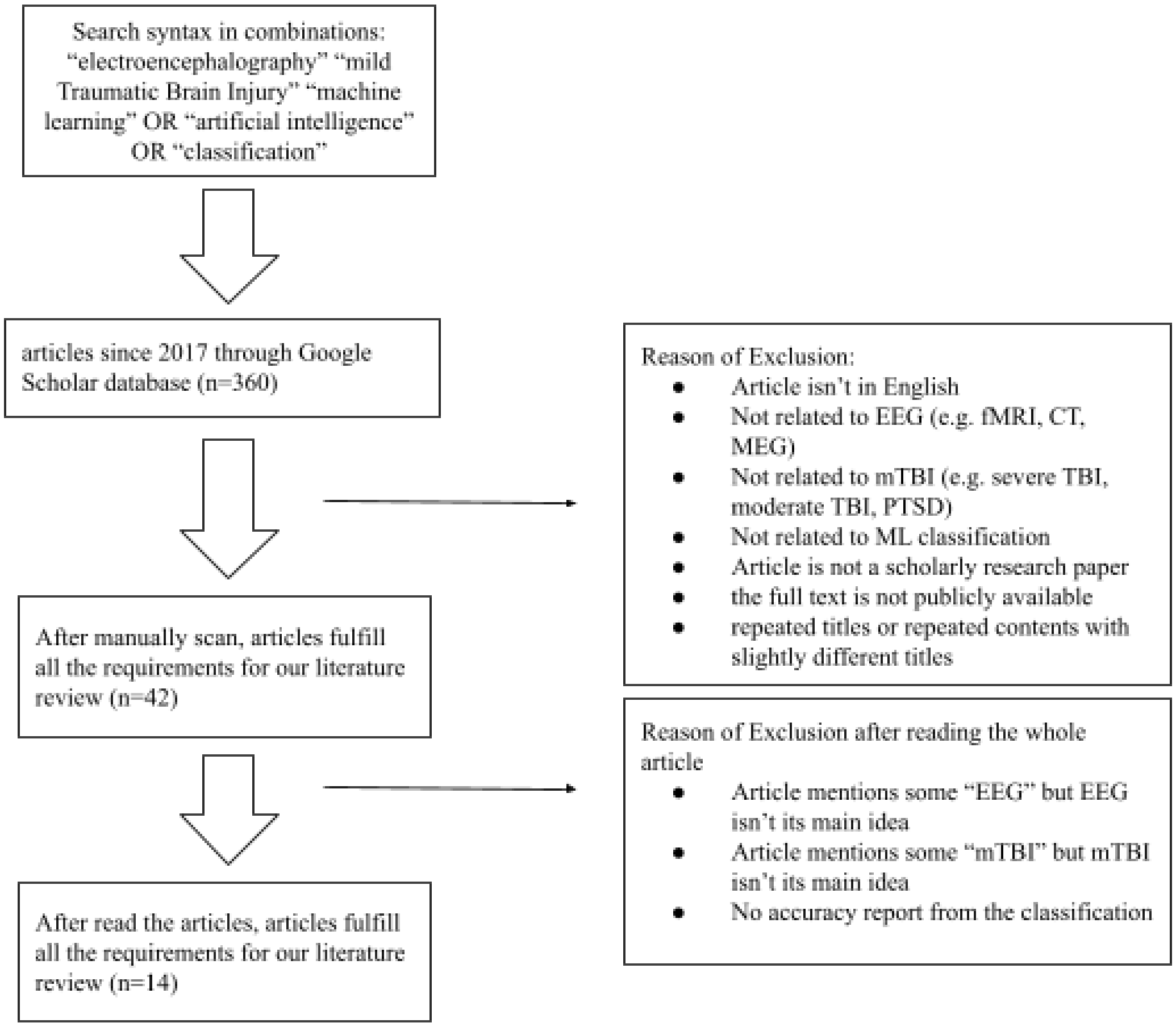

Figure 1. Search report flow chart describing the survey procedure, research keywords, and elimination reasons.

# III. Results

We conducted a rigorous review of the 14 most cutting-edge studies from 2017 to the present. In Table 1, we presented each study's objective and future impact. As the table shows, each of them has different objectives and impacts. This survey will review each article in five components: EEG data types, EEG data processing methods, feature extraction methods, features, and machine learning classifiers. Each component will be explained and analyzed in the sections down below.

| **Publications** | Time | Challenge | Impact |
|---|---|---|---|
| Lewine et al.[56] | 2019 | find biomarkers for mTBI in chronic periods | indicate that quantitative EEG methods can be useful in the identification, classification, and tracking of individual subjects with mTBI |
| Jafarlou et al.[57] | 2020 | investigate various machine learning approaches on a unique 24-hour recording dataset of a mouse TBI model | Results were promising with an accuracy of 80-90% with appropriate features and parameters using a small number of objects |
| Lai et al.[58] | 2020 | To offer an instant brain condition classification system for acute mTBI | Offers an opportunity to quickly and accurately rest state EEG data. |
| Dhillon et al.[59] | 2021 | to create a compact and portable system for early detection of TBI that can be used in real-life rather than in theory | Deployed on a Raspberry Pi 4 showing the same prediction metrics as a general computer |
| McNerney et al.[60] | 2017 | Identify mTBI using not only a symptom questionnaire but also EEG data | 1) The combination of EEG and symptom questionnaires is more accurately classified mTBi than questionnaires alone 2) EEG can be used to reduce the subjectivity of the remove-from-play decision. 3) The boosting method in ML is a powerful tool for enhancing classification accuracy |
| Vishwanath et al.[61] | 2020 | To determine which ML algorithm is the most effective in classifying mTBI data | Shows that CNN has more potential in classifying mTBI data while also proving that other ML algorithms while giving pretty high accuracy, can't compare to the flexibility of CNN. |
| Thanjavur et al.[62] | 2021 | Find an objective, clinically- accepted, brain-based approach for determining whether an athlete has suffered a concussion. | This shows that RNN is an effective method of classifying concussions from EEG data. |
| Bazarian et al.[63] | 2021 | Utilize the Concussion index as an objective and effective method in identifying mTBI | An objective, reliable indicator of the presence and severity of concussive brain injury and the readiness for the return to activity has the potential to reduce concussion-related disability. |
| Lee et al. [64] | 2020 | develop a classifier on whether the individual is concussed or not based on EEG | show  that SVM has the potential to provide an alternative solution to concussion analysis |
| Boshra et al.[65] | 2019 | To use mTBI detection based on EEG/ERP analysis in clinical assessment | 1) show that CNN enabled an end-to-end solution with minimal feature-engineering 2) supported the hypothesis that single-trials offer a more effective method of assessing EEG/ERP data. 3) first report of ML-based EEG/ERP analysis in acute/post-acute concussion assessment. |
| Sutandi et al.[66] | 2020 | Find a way for a low-cost and fast method for detection of TBI | first system deployed a portable, low-cost, fast detection system like Raspberry Pi, to automatically classify for TBI detection |
| Rosas [67] | 2019 | To see if there were any changes in classification accuracy between visits. | The main purpose is to provide a further understanding in the research of functional connectivity as a reliable biomarker to be incorporated into the current clinical mTBI assessment protocol. The study was novel in this field by applying three popular connectivity estimators; PLV, iPLV, and AEC applied to resting-state EEG analysis while also assessing intra and cross-frequency coupling interactions. |
| Lai et al. [68] | 2021 | To see if it is possible to accurately classify TBI into three categories. | Proposed novel CNN ECOC-SVM voting ensembles architecture that uses pre-processed EEG presents high-performance measures |
| Buchanan et al.[69] | 2021 | The paper tries to offer a motor solution that caused EEG classification, specifically to diffuse axonal injury in mTBI | The first demonstration of a potential biomarker for comorbid PCS + CP from resting state EEG. |

Table 1. Survey studies summary

## 3.1 EEG Data Types

Researchers use many different methods to acquire EEG data in different settings. Our survey found that there are three main EEG recording methods: EEG with stimulants, resting-state EEG, and sleep-wake EEG. All four publications that use mice EEG data also use the sleep-wake recording method as the other two methods are impossible to implement on mice. The recording method with human EEG data is split between resting-state EEG and EEG with stimulants. In Table 1, 62% of the publications use resting-state EEG, 30% use sleep-wake EEG, and 8% use EEG with stimulants. There are a total of 13 articles in table 1. The publication Lee et al.[64] is excluded as the author does not include any specific type of EEG data in the description. As a result, we conclude that the resting state EEG is the most common EEG recording method used for mild TBI classification on human EEG data.

Resting-state EEG recently has been proved to have significant concussion diagnostic meaning [44]. The reasoning behind this widespread use of resting-state EEG is due to its ability to mimic the real situation as it requires a less active response from the patients. Although EEG with stimulants can often record a more robust result, it relies highly on patients' cognitive ability, such as attention or language comprehension [8]. Furthermore, people with brain injury, having altered brain function, might have problems performing these tasks. In addition, these stimulants can disturb the sleep-wake rhythm and affect the recovery process [9,10]. Thus, the resting-state EEG is more applicable in real life.

Table 2 shows two types of EEG contributors: humans and mice. Of course, human EEG data is always the most desirable one, but mice EEG data often shows similar behavioral deficits and pathology after the fluid percussion injury (FPI). In our survey, 71% of the publications use human EEG data, while 29% use mice EEG data. This finding concludes that human EEG is the more common EEG data used by researchers for mild TBI classification than mice EEG. It also shows an increasing availability of human EEG data and increasing attention and support from the public.

| EEG data recording methods | Publication(s) | EEG data Contributors | Publication(s) |
|---|---|---|---|
| resting state EEG (RS) | Lewine et al. [56],<br>Lai et al. (2020) [58],<br>McNerney et al. [60],<br>Thanjavur et al. [62],<br>Bazarian et al. [63],<br>Rosas [67],<br>Lai et al. (2021) [68],<br>Buchanan et al. [], | Human (H) | Lewine et al. [56],<br>Lai et al. (2020) [58],<br>McNerney et al. [60],<br>Thanjavur et al. [62],<br>Bazarian et al. [63],<br>Lee et al. [64],<br>Boshra et al. [65],<br>Rosas [67],<br>Lai et al. (2021) [68],<br>Buchanan et al. [] |
| Sleep-wake EEG (SW) | Jafarlou et al. [57],<br>Dhillon et al. [59],<br>Vishwanath et al. [61],<br>Sutandi et al. [66], | Mice (M) | Jafarlou et al. [57],<br>Dhillon et al. [59],<br>Vishwanath et al. [61],<br>Sutandi et al.[66] |
| EEG with stimulants (stim) | Boshra et al. [65], | | |

Table 2. EEG data recording methods and EEG data contributors

## 3.2 Data Filtering Methods

EEG also possesses poor spatial resolution [15] due to their placement, meaning that neurons buried within the sulci or deep brain structures have far less contribution to the EEG signal. Additionally, since EEG generally has a low signal-to-noise ratio [16], the brain is often buried under many sources of noise of similar to or greater amplitude called “artifacts.” Consequently, various techniques and methods must be employed to prevent, remove, or lessen the impact of noise.

In the papers we collected, there were cases where researchers tested the necessity of EEG data preprocessing using raw EEG data as the input. However, the majority of studies utilized some filtering methods where epochs of the EEG data were sorted into different frequency bands: delta (1.0–3.5 Hz), theta (4.0–7.5 Hz), alpha (8–12 Hz), beta (12.5–25 Hz), high beta (25.5–30 Hz), and gamma (30–50 Hz)[56]. Although, the number of bands varies between studies. Several studies also employ some artifact removal methods via independent component analysis(ICA) methods. Additionally, two studies (Lewine et al.[56], Lai et al.(2020)[58]) utilize visual inspection, but only as a final screening and not as the primary artifact removal method. This is due to the arduous nature of artifact rejection or manual rejection, which may require a person dedicated to eliminating artifacts visually one by one in an EEG [37].

In Figure 2, a visual representation of the collected studies is displayed, with most studies utilizing bandpass filtering as their preferred method of data preprocessing. An outlier within the studies employing bandpass filtering is Buchanan et al. [69], which removed artifacts and then passed the processed EEG data through the bandpass filter. Another outlier is Lewine et al. [56], which processed data using Neuroguide, a software designed to handle various tasks, including filtering and artifact rejection. As mentioned before, Thanjavur et al. [62] entirely skip the data preprocessing stage and directly utilizes raw EEG data as the input for the machine learning algorithm. Lee [64] also does not mention any preprocessing methods, and thus it is excluded in the figure. Besides those outliers, most studies move linearly. First, bypass raw EEG data through bandpass filters, then separate them into different frequencies. Some papers stop at this step and test the classification accuracies of each band; however, five studies continue by downsampling the data and removing any artifacts either via a computer program or manually. Finally, Lai et al. [58] utilize the bootstrap method to process the data and further consolidate the EEG data.

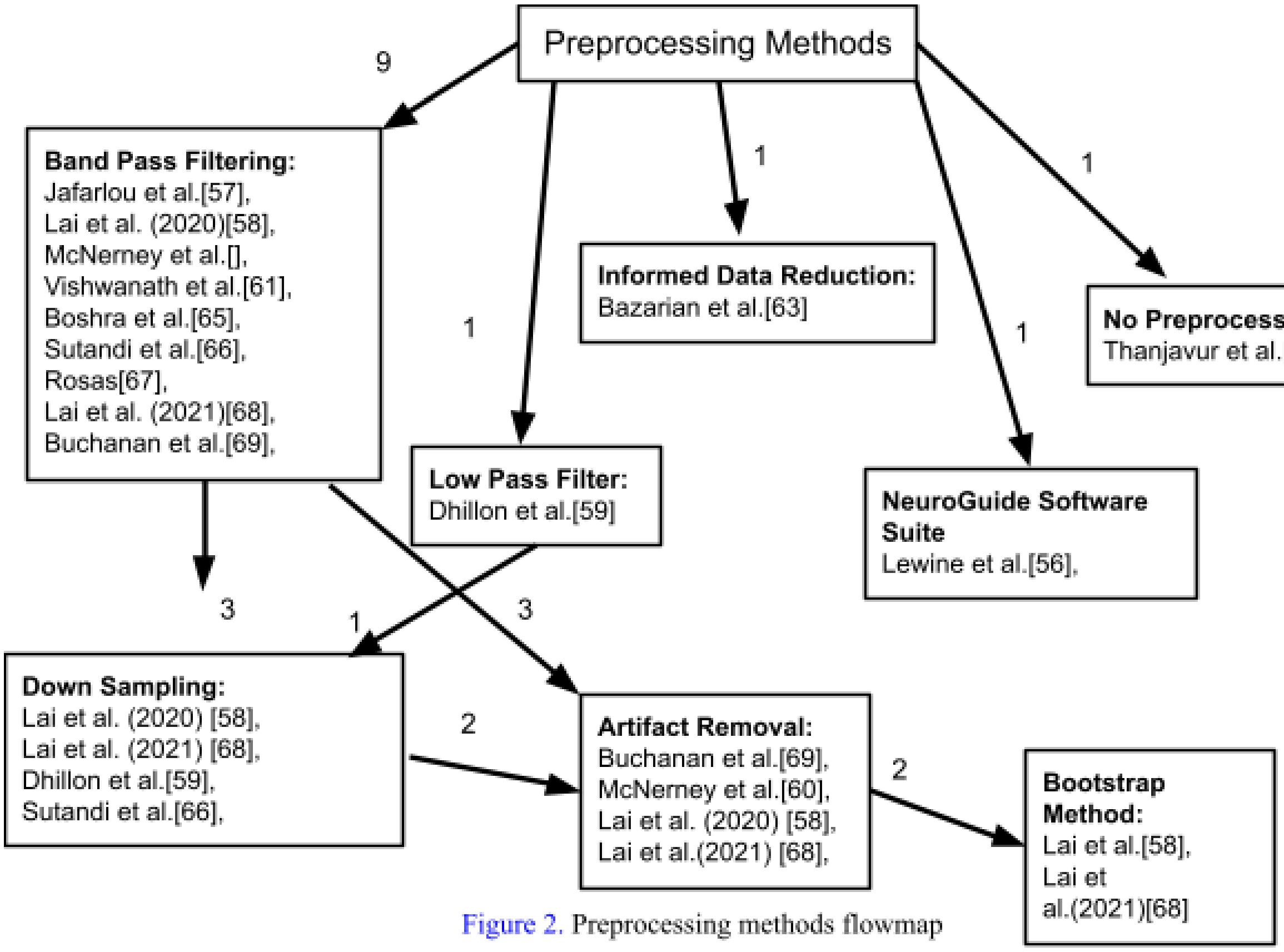

Figure 2. Preprocessing methods flowmap

### 3.3 Machine Learning Features

EEG has many data points, and researchers need to pick those points that have a significant difference between mTBI patients and healthy people to attain an optimal classification accuracy.

As mentioned in the introduction, mTBI patients often vary significantly in the duration after injury, the injury location, severity of the injury, and the patients themselves. Therefore, classification models that featured only one variable often performed poorly. This is because many features often overlap between mTBI and healthy people. For example, Lewine's model only resulted in an optimized accuracy of 58% with a univariate approach [56]. Thus, a model featured with multiple qEEG indicators with machine learning is necessary.

As shown in Table 3, many studies in our survey focused on spectral power: delta, theta, alpha, beta, and gamma. These spectral powers then use different mathematical operations to attain other EEG variables, including global relative powers and absolute powers, interhemispheric coherence, average power, power spectral density, spectral correlation coefficient, etc. The alpha and theta frequency bands are frequently used among the standard features. Numerous studies prove that there is a significant increase in theta power and a significant decrease in alpha power [56, 61]. One thing needs to be pointed out. The neural networks can perform feature extraction automatically. Thus, they do not need any human-input features. The qEEG indicators are used as feature vectors in rule-based models, such as support vector machine (SVM), K-nearest neighbor (KNN), and decision tree (DT).

McNerney [60] uses a very interesting combination between EEG features and non-EEG features in all the publications. It uses the power spectral bands as the EEG input while using the traditional symptom questionnaire as the non-EEG input. The questions are about the loss of consciousness, headache, nausea, vomiting, sensitivity to light, sensitivity to sound, confusion, and memory dysfunction. The patients then will rate their symptom severity in the interval of 0 to 6. This setting shows promising results with the average AUC values of 52% of the EEG-only trial compared to the 91% of the EEG plus symptom trial. Moreover, as the symptom questionnaire eliminates many possibilities, the study only needs three electrodes to collect the EEG, significantly reducing the demand for computational power and storage. This study shows promises of utilizing the EEG classification system on a real-life clinical basis. Furthermore, it shows the importance of using EEG with other variables to attain the desired result.

Although spectral powers are the most common EEG indicators, many potential helpful features are also. In Lee's study [64], the model is featured with both spectral powers and entropy. According to Lee, the model can return an 99.97% accuracy on mTBI classification. This result is shocking. We recommend that future researchers explore other non-spectral power features and integrate the most effective one into the model. According to Lee's result, this shows very successful results.

| Publications | Features |
|---|---|
| Lewine et al.[56] | global relative alpha & theta, global absolute high beta, interhemispheric coherence, global phase shift in beta |
| Vishwanath et al. | average power for rule-based models; automatically feature extraction for CNN |
| Lai et al. (2020) | LSTM automatically extract features |
| Dhillon et al.[59] | average power, PSD, average theta: alpha power ratio |
| McNerney et al.[60] | delta, theta, alpha, beta, sigma, and gamma bands (EEG variables) yes/no answers and the average of the numerical intensities of the seven symptoms questions (non-EEG variables) |
| Vishwanath et al. | average power, alpha: theta power ratio |
| Thanjavur et al.[62] | LSTM automatically extract features |
| Bazarian et al.[63] | total power, alpha band absolute asymmetry, beta band interhemispheric coherence |
| Lee et al. [64] | spectral powers, entropy |
| Boshra et al.[65] | CNN automatically extract features |
| Sutandi et al.[66] | CNN automatically extract features |
| Rosas[67] | Phase Locking Values (PLV), Imaginary PLV, and amplitude envelope correlation (AEC) |
| Lai et al. (2021) | CNN automatically extract features |
| Buchanan et al.[69] | absolute delta and theta, whole-brain average |

Table 3 EEG Features

## 3.4 Feature Extraction Methods

EEG classification needs features. These features are usually hidden within the wavy data lines and need some unique technique to dig them out. This survey examined 14 studies, as shown in Table 4. The most common feature extraction method is discrete Fourier transform, fast Fourier transform, and Welch's method. Jafarlou [57] and Vishwanath [61] use Discrete Fourier Transform, and Lewine [56] and Lee [64] use Fast Fourier Transform. Lastly, Dhillon [59], McNerney [60], and Buchanan [69] both use Welch's method.

Our survey showed a correlation between the use of neural networks as feature extraction methods and high classification accuracy. Lai in 2020 [58] used both LSTM and in 2021 [68] used CNN to extract features automatically. The classification accuracy is, respectively, 100% and 99.76%. In another study, Thanjavur [62] also used the LSTM as automatic feature extraction and attained an AUC of 97.1%. These neural network layers automatically remember important info from each time step and find correlations between them. These correlations are then extracted by backpropagation through time and stored as activations. Using neural networks as feature extraction seems to be a promising method. Because of its automatic feature extraction, neural networks can level up the classification time, decrease the need for specialized people, and make the model fit in all circumstances.

| Publications | Feature Extraction Methods |
|---|---|
| Lewine et al.[56] | Fast Fourier Transform, cross-spectral analysis |
| Vishwanath et al.[57] | Discrete Fourier Transform, decibel normalization |
| Lai et al. (2020)[58] | LSTM |
| Dhillon et al.[59] | Discrete Fourier Transform |
| McNerney et al.[60] | Discrete Fourier Transform, symptom questionnaire (non-EEG variables) |
| Vishwanath et al.[61] | Discrete Fourier Transform, decibel normalization |
| Thanjavur et al.[62] | LSTM |
| Bazarian et al.[63] | Concussion Index Algorithm |
| Lee et al. [64] | Fast Fourier Transform, Wavelet Analysis, Shannon Entropy |
| Boshra et al.[65] | visual inspection, ICA decomposition |
| Sutandi et al.[66] | Feature Extraction layer in CNN |
| Rosas[67] | Minimum Redundancy Maximum Relevance |
| Lai et al. (2021)[68] | CNN |
| Buchanan et al.[69] | Discrete Fourier Transform |

Table 4. Feature Extraction Method Table

## 3.5 Machine Learning Classifiers

In our paper, these machine learning approaches were the most commonly used. However, It must be noted that many more machine learning algorithms can be applied to this application. Figure 3 illustrates the prevalence of each machine learning method in the collected papers. In the table, we can see a clear trend towards support vector machines (SVM), appearing a total of eight times, and convolutional neural networks, which appear seven times. The prevalence of these two algorithms signals a clear advantage in classification accuracy and ease of use. Notably, Lewine et al.[56], Vishwanath et al.[57], Dhillon et al.[59], Vishwanath et al.[61] test multiple machine learning algorithms, while most test a single method.

In Figure 3, a clear pattern can be seen with SVM consistently having the highest accuracy, some even as high as 100% in Lai et al. [58]. Boosting comes in second with the highest classification accuracy at 98% in Dhillon et al.[59], which may be attributed to its scalability that allows parallel and distributed computing and makes learning and model exploration faster. RNN was used only once in Thanjavur et al[62], however, it yielded promising results of 88.9% classification accuracy. Multilayer Perceptron and Naive Bayes performed comparatively worse than other previously mentioned methods. NN and GA performed better, accuracy-wise than MP and NN. However, It must be noted that many of these statistics were derived from limited EEG data sets, some as small as eleven total subjects, and as a result, accuracy may be skewed with different margins of error.

Convolutional Neural Networks (CNN) comes in third with the highest classification accuracy at 92.03% in Vishawanth et al. [57], although it is second in prevalence, appearing in over half of studies. As mentioned above, CNN is suited for image classification due to its framework, making it more suitable for applications such as pneumonia detection[70]. Lai et al. (2021)[68] utilize both CNN and SVM. However, they only employ CNN for its automatic feature extraction functions [46], another reason why a few papers solely use CNN.

A reason why SVM may have been used most often could stem from its low computational complexity, the small number of variables, and solid mathematical theoretical basis. SVM has a lower computational complexity when compared to other typical ML classifiers such as multi-layer perceptron (MLP) and KNN [53], [54]. SVM also has fewer parameters than other deep learning methods; however, these parameters must be changed manually and greatly influence classification accuracy and results [50]. It must be noted that researchers are working towards removing the parameter adjusting process of neural networks. However, CNN has its advantages which have been previously discussed. Additionally, CNN may have been favored due to its ability to learn features from local receptive fields and detect abstract features through convolutional layer repetition, making them suitable for complex EEG recognition problems[55].

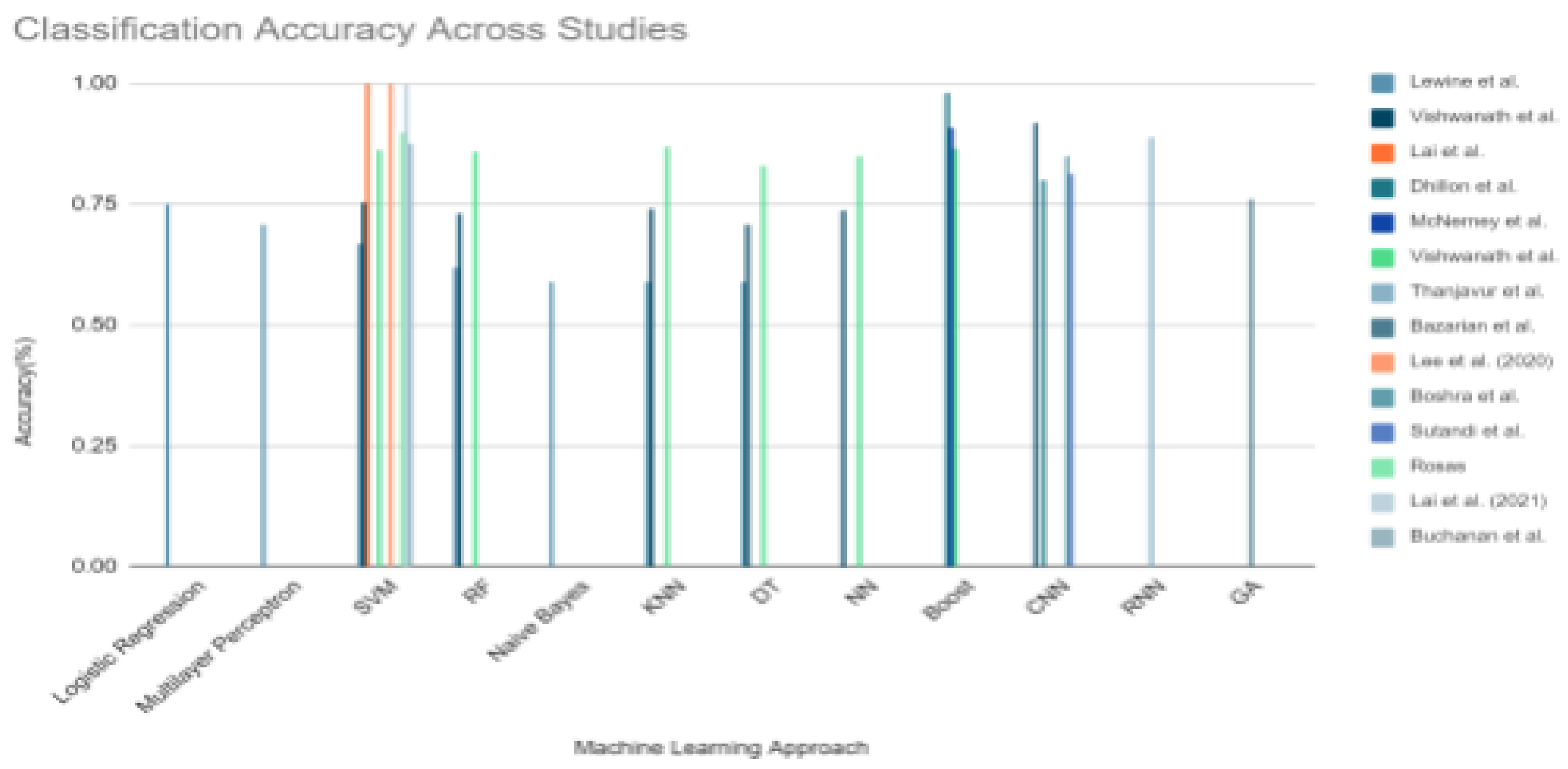

Figure 3. Accuracies Across Studies

# IV. Discussion

| classifier | accuracy | extraction | EEG type | data size | preprocess | Pub. | features |
|---|---|---|---|---|---|---|---|
| SVM | 100% | LSTM | RS, H | 64 EEG Channel | BP filter, DS, AR, Boot. | Lai (2020)[58] | ---- |
| | 99.97 % | PSA, WA, SE | unknown, H | 40 total subjects | ------- | Lee (2020) [64] | SP, entropy |
| | 99.76 % | CNN | RS, H | 64 EEG Channel; 350 data points for each channel | BP filter, DS, AR, Boot. | Lai (2021)[68] | ---- |
| | 90% | MRMR | RS, H | 24 channel; 300 data points for each channel | BP Filter | Rosas[67] | PLV, iPLV, AEC |
| | 87.60 % | WM | RS, H | 19 channel; 180 data points for each channel | BP filter, AR | Buchanan [69] | AP(δ, θ), WBA |
| | 86.15 % | DFT, DN | SW, M | 86400 data points | BP Filter | Vishwanat [57] | aveP, α:θ ratio |
| | 75.60 % | DFT, DN | SW, M | 86400 data points | BP Filter | Vishwanath [61] | aveP |
| | 67% | FFT, CSA | RS, H | 300 - 420 data points; 7130 - 28365 data points | NeuroGuide | Lewine[56] | RP(α, θ) AP(high β), IC, GPS in β |
| Boost | 98% | DN, WM | SW, M | 1 channel; 86400 data points | LP Filter, DS | Dhillon[59] | aveP, α:θ ratio |
| | 91% | WM, SQ | RS, H | 60 datapoints for each subject | BP filter, AR | McNerney [60] | δ, θ, α, β, σ, ɣ, SQ |
| | 86.48 % | DFT, DN | SW, M | 86400 data points | BP Filter | Vishwanat [57] | aveP, α:θ ratio |
| CNN | 92.03 % | DFT, DN | SW, M | 86400 data points | BP Filter | Vishwanath [61] | ----- |
| | 80% | CNN | SW, M | 1 channel; 86400 data points | LP Filter, DS | Dhillon[59] | ----- |
| | 85% | CNN | Stimu, H | 64 channel | BP Filter | Boshra[65] | ----- |
| | 81.50 % | CNN | SW, M | 1 channel; 86400 data points; or 60 data points for each subject | BP Filter, DS | Sutandi[66] | ----- |
| RNN | 88.90 % | RNN | RS, H | 64 channel; 300 data points for each subject | raw data | Thanjavur [62] | ----- |
| KNN | 87% | DFT, DN | SW, M | 86400 data points | BP Filter | Vishwanat [57] | aveP, α:θ ratio |
| | 74.30 % | DFT, DN | SW, M | 86400 data points | BP Filter | Vishwanath [61] | aveP |
| | 59% | FFT, CSA | RS, H | 300 - 420 data points(EEG Recording Length); 7130 - 28365 data points(total # of cleaned artifacts) | NeuroGuide | Lewine[56] | RP(α, θ) AP(high β), IC, GPS in β |
| RF | 86% | DFT, DN | SW, M | 86400 data points | BP Filter | Vishwanat [57] | aveP, α:θ ratio |
| | 75.60 % | DFT, DN | SW, M | 86400 data points | BP Filter | Vishwanath [61] | aveP |
| | 62% | FFT, CSA | RS, H | 300 - 420 data points; 7130 - 28365 data points(total # of cleaned artifacts x # of subjects) | NeuroGuide | Lewine[56] | RP(α, θ) AP(high β), IC, GPS in β |
| NN | 84.85 % | DFT, DN | SW, M | 86400 data points | BP Filter | Vishwanat [57] | aveP, α:θ ratio |
| | 73.80 % | DFT, DN | SW, M | 86400 data points | BP Filter | Vishwanath [61] | aveP |
| DT | 82.90 % | DFT, DN | SW, M | 86400 data points | BP Filter | Vishwanat [57] | aveP, α:θ ratio |
| | 71% | DFT, DN | SW, M | 86400 data points | BP Filter | Vishwanath [61] | aveP |
| | 59% | FFT, CSA | RS, H | 300 - 420 data points; 7130 - 28365 data points(total # of cleaned artifacts x # of subjects) | NeuroGuide | Lewine[56] | RP(α, θ) AP(high β), IC, GPS in β |
| GA | 76% | CIA | RS, H | 600 data points for each subject; or 60 - 120 data points for each subject(total # of cleaned artifacts x # of subjects) | Data Reduction | Bazarian[63] | |
| *LR | 75% | FFT, CSA | RS, H | 300 - 420 data points(EEG Recording Length); 7130 - 28365 data points (total # of cleaned artifacts x # of subjects) | NeuroGuide | Lewine[56] | RP(α, θ) AP(high β), IC, GPS in β |
| *MP | 71% | FFT, CSA | RS, H | 300 - 420 data points(EEG Recording Length); 7130 - 28365 data points (total # of cleaned artifacts x # of subjects) | NeuroGuide | Lewine[56] | RP(α, θ) AP(high β), IC, GPS in β |
| NB | 59% | FFT, CSA | RS, H | 300 - 420 data points(EEG Recording Length); 7130 - 28365 data points(total # of cleaned artifacts x # of subjects) | NeuroGuide | Lewine[56] | RP(α, θ) AP(high β), IC, GPS in β |

Table 5 ML Process Summary of Research Papers

The scope of our survey may be limited to a relative niche sector of machine learning applied to mTBI. However, its implications for soldiers, constructors, athletes, drivers, children, elders, and many more people who suffer or are at risk for mTBI may signal a change in the fundamental way we treat and diagnose brain injuries. More specifically, the studies we collected identified a clear trend toward SVM as an ideal classifier in this application. Deep learning methods, such as CNN and RNN, although not as accurate in this application, may prove more versatile due to their built-in feature extraction capabilities, which SVM and many other rule-based ML algorithms unfortunately lack. Some studies [58], [68] use combined classifiers in the survey: a deep learning algorithm for automatic extracting features and a rule-based algorithm for classification. In this way, the result seems promising: 100% [58] and 99.76% [68]. In this combination, both deep learning and rule-based machine learning use their strengths. Rule-based algorithms provide a clear cut for classification but often need human-input features. The human-input features oftentimes are both time-consuming and not accurate, as shown in most of Lewine's rule-based classifiers [56]. On the other hand, deep learning, such as CNN, shows a mediocre accuracy (average of 85%), but are excellent feature extractors.

Although many conventional machine learning algorithms, such as Decision Tree, Multilayer Perceptron, and Logistic Regression show a low accuracy, these methods are still worth of research in the future. In fact, Boosting Classifiers, each of which is often a collection of weak performance algorithms, shows a stable high accuracy among the three studies: [Dhillon 59], 91% [McNerney 60], and 86% [Lewine 56]. This result suggests that grouping these algorithms together in a collection may provide new approaches for future studies though these individual classifiers provide low results.

Another thing has to be pointed out. All the processes can contribute to the final accuracy rather than solely the classifiers. Our survey shows a great diversity of choices in EEG data recording methods, EEG data contributors, data processing methods, feature extraction methods, feature representations, and machine learning classifiers. As Table 1 shows, each study has a slightly different aim, and thus their choices of the process can be different. Some of the studies aim for a high accuracy to prove the effectiveness of EEG. Therefore, they will spend more time cleaning and filtering the data and finding the most practical and effective features that suit this particular dataset, resulting in high accuracy, while the re-testability with different data might be lower as mTBI EEG data is known for its variability. Others consider the testability problem and use algorithms that do not need human feature inputs. The system might also require more computational power and storage space, which is not always available in a portable device setting. Finally, others aim for real application in clinical situations. Therefore, they might choose the feature extraction methods and machine learning classifiers that best suit the clinical situation and leave some accuracy for speed, price, and portability. This survey summarized the most cutting-edge studies after 2017 and pointed out the general trend of the studies. We suggest future researchers consider all five factors and choose the best aligned with their objectives.

It is crucial to understand the limited accessibility of publicly available data. Table 6 lists the datasets available to the public. However, many need to be requested from the author or relevant sources. Much of the data used in our studies employed exclusive data collected from a few locations and subjects. Moreover, many publicly accessible datasets such as the UCI database and OpenNeuro lack substantial abnormal mTBI datasets. Vishwanath et al.[61] utilized a dataset of nine mice, four controls, and five concusses, which may have caused results to be skewed and inaccurate. However, there has been a movement towards standardizing assessment and analytic tools, facilitating the overarching goal of distributed data collection and data mining through the work of Cavanagh et al. (2017)[52] and more. Simply, open-source EEG databases lack quantity and lead to poor generalizability across literature due to the high variability of subjects, testing methods, etc. While there was an attempt to validate the accuracy of the studies, such as Lee(2020)[64], the outdated use of MATLAB code made it difficult to reproduce similar results in Python. It is also worth noting that more advanced machine learning methods integrate both advantages of SVM and Deep Learning, such as Geometric Unified Learning (https://www.sbir.gov/node/2080689) developed by Dasion, which can reach 96% - 100% accuracy in mTBI predictions, are needed in future development.

Finally, using machine learning on EEG data may save hundreds of lives. In Dhillon et al.[59], a small portable raspberry pi-based mTBI detector was created that has the potential of being a cheap, safe, and accurate means at identifying mTBI on scene. In the future, CT scans may become obsolete, and smaller, lighter devices using EEG data may be favored especially for mTBI detections. There are still many things to discover and many problems to be solved, but we are close to using advanced ML in healthcare.

| Publications | Data | Code |
|---|---|---|
| Dhillon et al.[59] | Available upon request | Not Listed |
| Lee et al. (2020)[64] | Concussed data control data | Github |
| Boshra et al.[65] | Available upon request | Github |
| Sutandi et al.[66] | Not Listed | Github |
| OpenNeuro | EEG Data | N/A |
| UCI | EEG Data | N/A |
| Cavanagh | mTBI Data | N/A |

Table 6. Available published datasets and code repositories

# V. Conclusion

In conclusion, this survey analyzed the 14 most cutting-edge machine learning research papers from 2017 to the present in terms of 1) EEG data types, 2) data preprocessing methods, 3) machine learning features, 4) feature extraction methods, and 5) machine learning classifiers. The most common EEG data type was human resting-state EEG, with most studies using filters to clean the data. Power spectral was used most heavily, especially the alpha and theta power, while other non-power spectral features, such as entropy, also showed great potential. Of the many feature extraction methods used by the collected studies, Fourier transform was used most frequently. However, neural networks were also used heavily as their automatic feature extraction capabilities generally returned equally high accuracies. Regarding ML algorithms, SVM was our survey's most common classifier and consistently had the highest classification accuracy among the other methods. Most of the studies in our survey are still not in the stage of clinical use. In the future, we believe that there will be an optimal combination of different machine learning approaches such as Geometric Unified Learning, which take advantage of various different ML approaches, that can achieve a fast, cheap, and portable mTBI detection for people in the world who suffer from mTBI.